\documentclass{IEEEcsmag}
\usepackage[colorlinks,urlcolor=blue,linkcolor=blue,citecolor=blue]{hyperref}
\expandafter\def\expandafter\UrlBreaks\expandafter{\UrlBreaks\do\/\do\*\do\-\do\~\do\'\do\"\do\-}
\usepackage{upmath,color}

\jvol{XX}
\jnum{XX}
\paper{8}
\jmonth{Month}
\jname{Publication Name}
\jtitle{Publication Title}
\pubyear{2024}

\usepackage{wrapfig}
\usepackage[numbers]{natbib}
\newcommand{\etal}{\textit{et al.}}
\newcommand{\ie}{\textit{i.e.}}
\newcommand{\eg}{\textit{e.g.}}

\begin{document}

\sptitle{FEATURE ARTICLE: Garment modelling, Sewing pattern design, \\Surface flattening}

\title{PerfectTailor: Scale-Preserving 2D Pattern Adjustment
Driven by 3D Garment Editing}

\author{Anran Qi and Takeo Igarashi}
\affil{The University of Tokyo, Tokyo, 113-0033, Japan}

\markboth{FEATURE}{FEATURE}

\begin{abstract}\looseness-1 We address the problem of modifying a given well-designed 2D sewing pattern to accommodate garment edits in the 3D space. Existing methods usually adjust the sewing pattern by applying uniform flattening to the 3D garment. The problems are twofold: first, it ignores local scaling of the 2D sewing pattern such as shrinking ribs of cuffs; second, it does not respect the implicit design rules and conventions of the industry, such as the use of straight edges for simplicity and precision in sewing. To address those problems, we present a pattern adjustment method that considers the non-uniform local scaling of the 2D sewing pattern by utilizing the intrinsic scale matrix. In addition, we preserve the original boundary shape by an as-similar-as-possible geometric constraint when desirable. We build a prototype with a set of commonly used alteration operations and showcase the capability of our method via a number of alteration examples throughout the paper.
\end{abstract}
\maketitle

\chapteri{C}urrent garment alternation design is mostly centered around 2D sewing pattern space which involves numerous pattern editing operations to achieve the envisioned alterations of 3D garments. In practice, achieving the correct pattern adjustment not only necessitates specialized expertise in garment design but is also time-consuming. This is because designers need to justify
both the envisioned 3D geometric changes of the garment and the embedded intrinsic design, such as smocking, elastic threading design, and \emph{etc}. 

\label{intro}
\begin{figure}[t]
    \centering
    \includegraphics[width=0.95\linewidth]{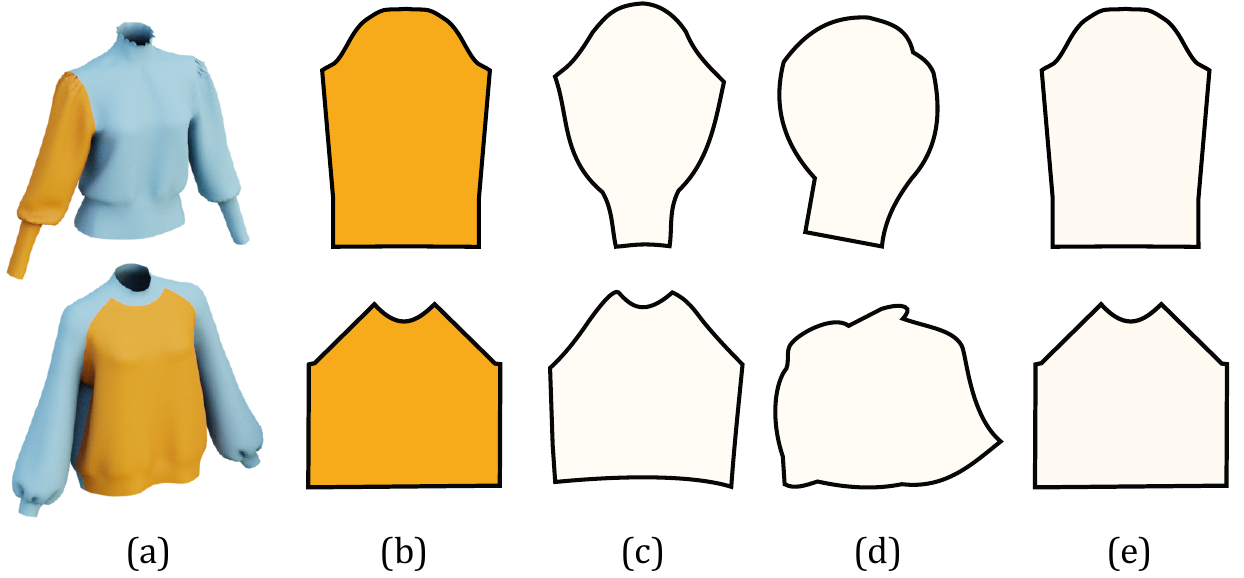}
    \vspace{-0.2cm}
    \caption{(a) Two 3D garments with hard yellow indicating the corresponding part to (b) the original panel designs. Panels flattened by  (c) Sheffer \etal$^{11}$ (d) Igarashi \etal$^4$ and (e) our method.  (c) and (d) generate smaller panels than the original panel (top row) and irregular boundaries (bottom row). In both cases, our method can produce the same panel shapes as that of the original panels. }
    
    \label{motivation}
\end{figure}

\begin{figure*}[ht]
 \centering
  \includegraphics[width=0.99\textwidth]{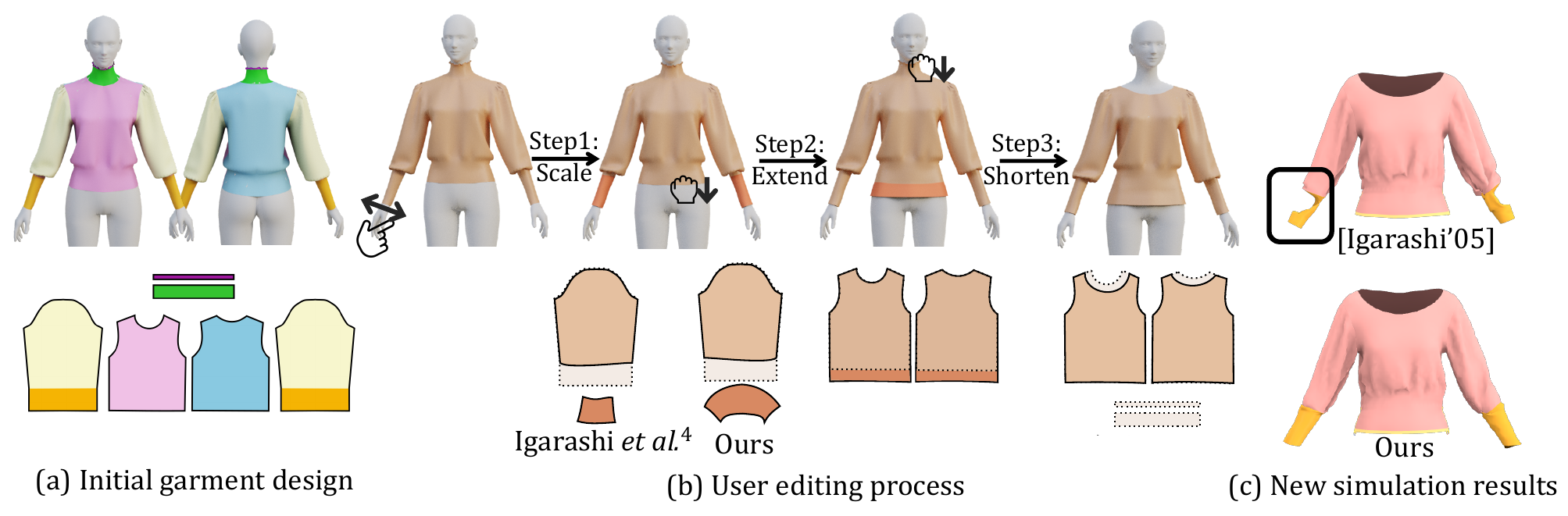}
  \caption{A user alteration example. \textmd{(a) A 3D garment model from the front and back view (top row) and its corresponding sewing pattern (bottom row). The matching colour indicates the garment and sewing pattern correspondence. It is worth mentioning that the bottom of the sleeve (hard yellow) fits tighter than the upper part (soft yellow) due to its elastic string design, while their corresponding pattern has a similar width. (b) Our function sets allow the user to alter the garment directly in 3D space based on their preference and our pattern adjustment method updates the pattern accordingly. Top row: the 3D garment edits by the user; Bottom row: the pattern adjustment results (Dash lines illustrate the original panel shape). For Step 1: scale operation, we show the pattern adjustment results of Igarashi \etal$^4$ which is a naive geometric surface parametrization technique and ours. Igarashi \etal$^4$ generates a much smaller panel due to directly flattening the surface. Thus unlike us, it is not able to preserve the original embedded design. Our system incorporates the left-right symmetry in garment design literature, \emph{i.e.,}
the user only needs to edit on a single side of the garment and our system automatically mirrors those edits across the other side. (c) The garment is simulated by the pattern from Igarashi \etal$^4$ (top row) and ours (bottom row). The smaller panel produced by Igarashi \etal$^4$ causes the  tearing of the sleeve highlighted in the black frame box.}}
  \label{teaser}
\end{figure*}

To speed up the design process and reduce the required expertise, researchers have proposed many powerful methods$^{1,2}$  to edit the garment directly in 3D space, and then automatically adjust the 2D pattern accordingly. Those methods usually assume the pre-existence of the sewing patterns. This matches the practice --- designers often start with an existing pattern to create either real or virtual outfits. As a well-designed sewing pattern has already undergone design and development processes, it can significantly streamline the design process for designers, saving time and resources in comparison to developing a new pattern from scratch. However, the existing methods use uniform flattening (geometric surface parametrization) to update the sewing pattern after 3D editing. This is suboptimal for the following two reasons. Firstly, it is grounded in the hypothesis that the local sizes of both 2D and 3D triangles in the sewing pattern and 3D garment design remain mostly constant.  For certain secondary textile designs like smocking and elastic thread design, the 3D triangle on the garment shrinks or expands due to embedded stitching force during the simulation and draping, which makes the local sizes of 2D and 3D triangles non-uniform.  Thus, uniformly flattening the 3D garment surface leads to misestimating the sewing panel size (see Figure~\ref{motivation} top row). Secondly, this technique usually produces panels with irregular boundaries. This neglects the implicit rules and conventions in the industry's design practices, such as straight edges, symmetric shapes and \emph{etc.} (see Figure~\ref{motivation} bottom row). 

Unlike previous work, we propose a pattern adjustment method specifically for synchronizing a well-designed 2D pattern according to the user’s edits in 3D space. Our method considers the non-uniform local scaling of the 2D sewing pattern and respects the implicit rules and conventions of the industry.   Specifically, our method memorizes the local intrinsic scale matrix between the 2D pattern and the 3D drape of the initial design and uses it when updating the 2D pattern to compensate for the user's edit in 3D space. In addition, we propose an as-similar-as-possible constraint to preserve the original panel boundary shape when the user modifies the entire panel. With our adjustment method, we develop a set of commonly-used editing operations to support garment alteration, such as scaling a specific part for fitting, extending and shortening the along the boundary, and cutting to achieve the desired shape (see Figure~\ref{teaser}) and demonstrate alteration results on a number of garments, showing its usability and generalizability. 

\begin{figure*}[ht]
    \centering
    \includegraphics[width=1.0\linewidth]{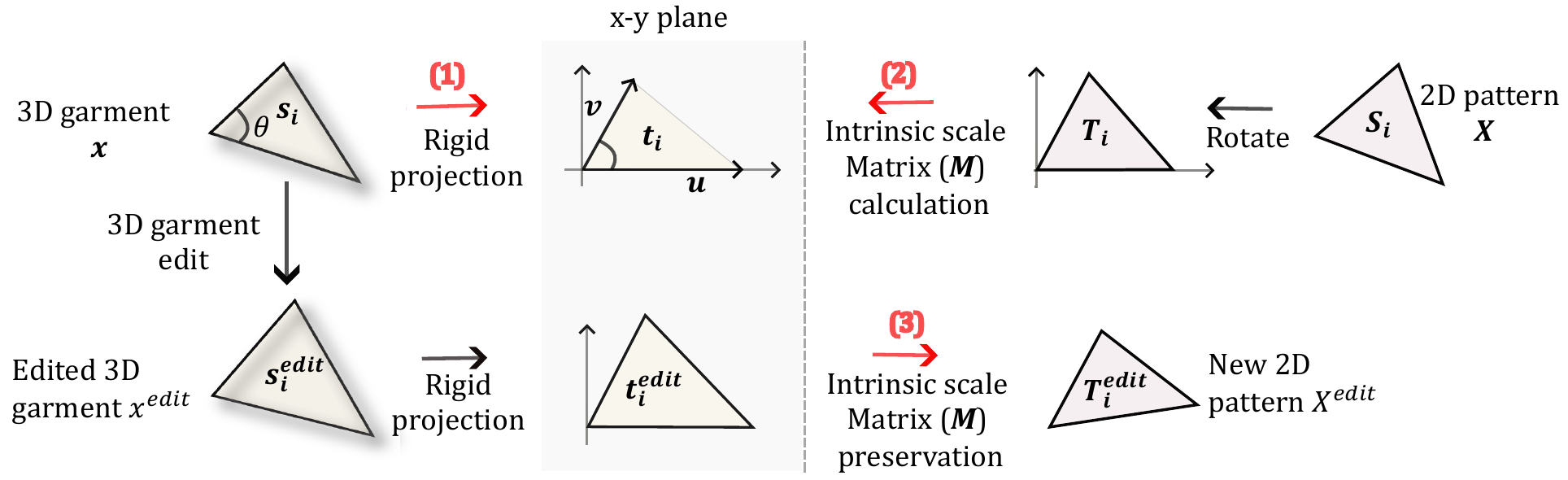}
     \vspace{-0.5cm}
    \caption{Computation flow of our flattening method. {\color{red} (1) (2)} and  {\color{red} (3)} indicate the calculation of Equation~\ref{eq0}, Equation~\ref{eq1} and Equation~\ref{eq2}, respectively.}
    \label{illu}
\end{figure*}

\section{RELATED WORK}
\subsection{Garment Design}
\label{R_GD}

Various computational garment design techniques offer tools to automate the adjustment of underlying 2D patterns. These methods contribute to accelerating the design workflow and minimizing the need for extensive expertise in the field.  Sensitive Couture$^{9}$ proposes a bidirectional interactive garment edit method by leveraging the fast simulation technique. It begins with well-designed sewing patterns and supports a subset of garment edits in 3D but is restricted to the dragging of vertices and edges.

Pietroni \etal $^{5}$ automatically generates the 2D sewing pattern from an input 3D shape by first creating the panel patch layout and then flattening the patch considering the anisotropic material property of the woven fabric.  Wolff \etal $^{10}$ optimizes the 3D rest shape of the garment to maximise the fit and comfort under a range of poses and body shapes. The corresponding sewing pattern is generated by Sharp \etal $^{6}$ which directly optimizes the distortion in the cutting and flattening process. These methods applies uniform flattening, thus cannot handle non-uniform local scaling of the sewing pattern and geometric constraints on boundary shapes.

Bartle \etal $^{1}$ proposes a fixed-point iteration method to optimize the pattern by minimizing the distortion between physical simulation results and the target draped garment so that it will generate the target garment geometry after simulation in the context of direct garment editing. The optimization accounts for pattern deformation caused by the physical forces during draping. However, physical simulation and optimization take time, requiring certain effort (\eg, parameter tuning) to make it work. Our method bypasses physical simulation by taking a purely geometric approach, which is faster and more stable. Combining our method with theirs can be an interesting future work.

\subsection{Surface Flattening}
\label{R_SF}
Surface flattening (\ie, parametrization) methods take a surface with disk topology and aim to optimize its 2D mapping based on the defined distortion measurement.  This is a fundamental and well-studied problem in computer graphics. We refer the reader to the survey$^{8}$
for a more complete background.  Here we review a few works that are commonly used in the garment design literature for completeness. Those works can be roughly classified into two categories: geometry-based methods$^{4, 11, 12, 13}$ and physics-based methods.$^{14, 15}$ 

Geometry-based methods formulate the surface flattening as a distortion minimization problem based on either vertex, edge, or face. Sheffer \etal$^{11}$ defines an angle preservation metric and a set of constraints on the angles to ensure the validity of the flat mesh. Then formulate it as a constrained minimization problem. As-rigid-as-possible methods use a local-global optimization approach to optimize an isometric distortion measurement defined based on a set of edges$^{12}$ or triangles.$^{4, 13}$

Physics-based methods use the elastic energy model to drive the deformation of each 2D triangle. McCartney \etal $^{14}$ proposes to use a relatively simple elastic model to calculate strain energy to measure the movement of vertices in each 2D triangle. Later, Wang \etal$^{15}$ converts the energy into force and utilizes it within the Lagrange equation to calculate the movement of each 2D vertex. 

Different from the aforementioned works which flatten the 3D surface from scratch, we focus on flattening the surface with an initial condition to keep the original design intention.

\section{METHOD}
Our key observation is the uniform flattening of the garment surface will dissipate the embedded initial design in the pattern. To this end, we propose to memorize the local scale matrix between the 2D pattern and the 3D drape of the initial design and use it when updating the 2D pattern to accommodate user editing on the 3D drape. Specifically, we seek a pattern adjustment method that transforms the original 2D pattern $X$ with 2D triangles $S$ to the new 2D pattern $X^{edit}$ with 2D triangles $T^{edit}$, which will follow the user's edit from the original 3D garment $x$ with 3D triangles $s$ to the edited 3D garment $x^{edit}$ with 3D triangle $s^{edit}$, whilst keeping the originally embedded design (see Figure~\ref{illu}).

\label{flatten}Motivated by Bartle \etal,$^{1}$ we aim at modelling the initially embedded local scaling as an
\emph{intrinsic scale matrix} and keep the matrix during the updating process. Since this matrix is invariant to rigid transformations, we can consider first projecting the 3D triangle $s_i$ of the 3D garment onto the 2D $x\mbox{--}y$ plane, denoted as $t_i$. Then rotate the $S_i$ to align the $t_i$, denoted as $T_i$, so that a designated edge corresponded in $t_i$ and $T_i$ is aligned with the $x$-axis (see Figure~\ref{illu}). Now we can formulate each triangle as 
\begin{equation}\label{eq0}
\begin{pmatrix}
\lvert u \rvert & \lvert v \rvert sin(\theta) \\
0 & \lvert v \rvert cos(\theta)
\end{pmatrix}
\end{equation}
where $u$ and $v$ are edge vectors and $\theta$ is the angle between them.
\noindent We model the intrinsic scale matrix as a 2D transformation $M$:
\begin{equation} \label{eq1}
\begin{split}
M_i = t_i T_i^{-1}
\end{split}
\end{equation}
At each time, the user edits the 3D garment, we update the 3D garment geometry based on well-established geometric rules and get the edited 3D triangle $s_i^{edit}$ with its 2D projection $t_i^{edit}$.

Recap that our goal is to keep the intrinsic scale matrix during the updating process. To do this, we seek to find the optimal 2D triangles $T^{edit}$ that when multiplied with this matrix will produce the target triangles $t^{edit}$:
\begin{equation} \label{eq2}
\begin{split}
t_i^{edit} = M_i T_i^{edit}
\end{split}
\end{equation}
Thus, we update the $T_i^{edit}$ as 
\begin{equation} \label{eq3}
\begin{split}
T_i^{edit} = M_i^{-1}t_i^{edit} = T_i t_i^{-1} t_i^{edit} 
\end{split}
\end{equation}

\noindent\textbf{Stitching} \label{stitching}After the updating stage, the new triangle has the desired property. However, each triangle is processed independently, so it is unlikely that they will not form a continuous 2D mesh pattern. We, therefore, need to stitch all the triangles together to form a valid pattern $S^{edit}$. This is a well-studied mesh parametrization problem that can be solved using geometry-based surface flattening methods. We use the as-rigid-as-possible surface flattening technique$^{4}$  to get our final results.

 After extensive deliberations with professional garment designers, we decided to preserve the original pattern shape, maintaining the integrity of the initial pattern when the user edit will affect the entire sewing panel equally. Therefore, we propose
 \setlength{\columnsep}{1pt}
\setlength{\intextsep}{1pt}
\begin{wrapfigure}{r}{0.2\textwidth}
    \centering
    \includegraphics[width=0.2\textwidth]{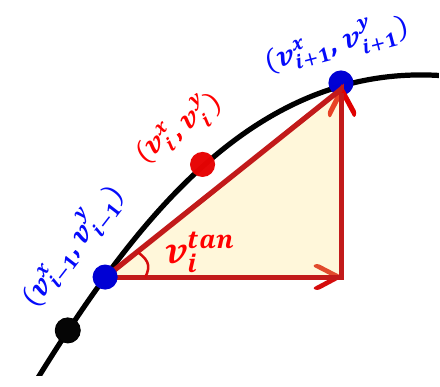}
\end{wrapfigure}
 an as-similar-as-possible constraint to preserve the discrete tangent of boundary vertices $v_i^{tan}$ expressed as $\frac{v^{x}_{(i+1)} -  v^{x}_{(i-1)}}{v^{y}_{(i+1)} -  v^{y}_{(i-1)}}$ in the stitching process (inset). We define final the quadratic error function as
\begin{equation}
\begin{split}
 &\arg\min\limits_{v' \in V} \sum_{(i,j) \in E} \Vert (v'_j - v'_i) - (v_j - v_i) \Vert^2\\  
 & + w_1\sum_{i \in C} \Vert (v'_i - C_i)\Vert^2 
 + w_2\sum_{i \in B} \Vert (v^{tan'}_i - B_i^{tan})\Vert^2
\end{split}
\end{equation}
where $v_i$ and $v_i^{'}$ are the vertex coordinate of the triangle in the original pattern ($T_i$) and the new targeted pattern ($T_i^{edit}$), respectively. $E$ is a set of edges,  $C$ is a set of fixed vertices, $C_i$  is the fixed vertex coordinate, $B$ is the set of boundary vertices and $B_i$ is the the boundary vertex coordinate. We fix the endpoints of an edge at the center of the
original pattern and use $w_1 = w_2 = 1000$ for the examples in this paper.

\section{SYSTEM OVERVIEW and IMPLEMENTATION}
\subsection{User Interface}
\begin{figure}[h!]
    \centering
    \includegraphics[width=1.0\linewidth]{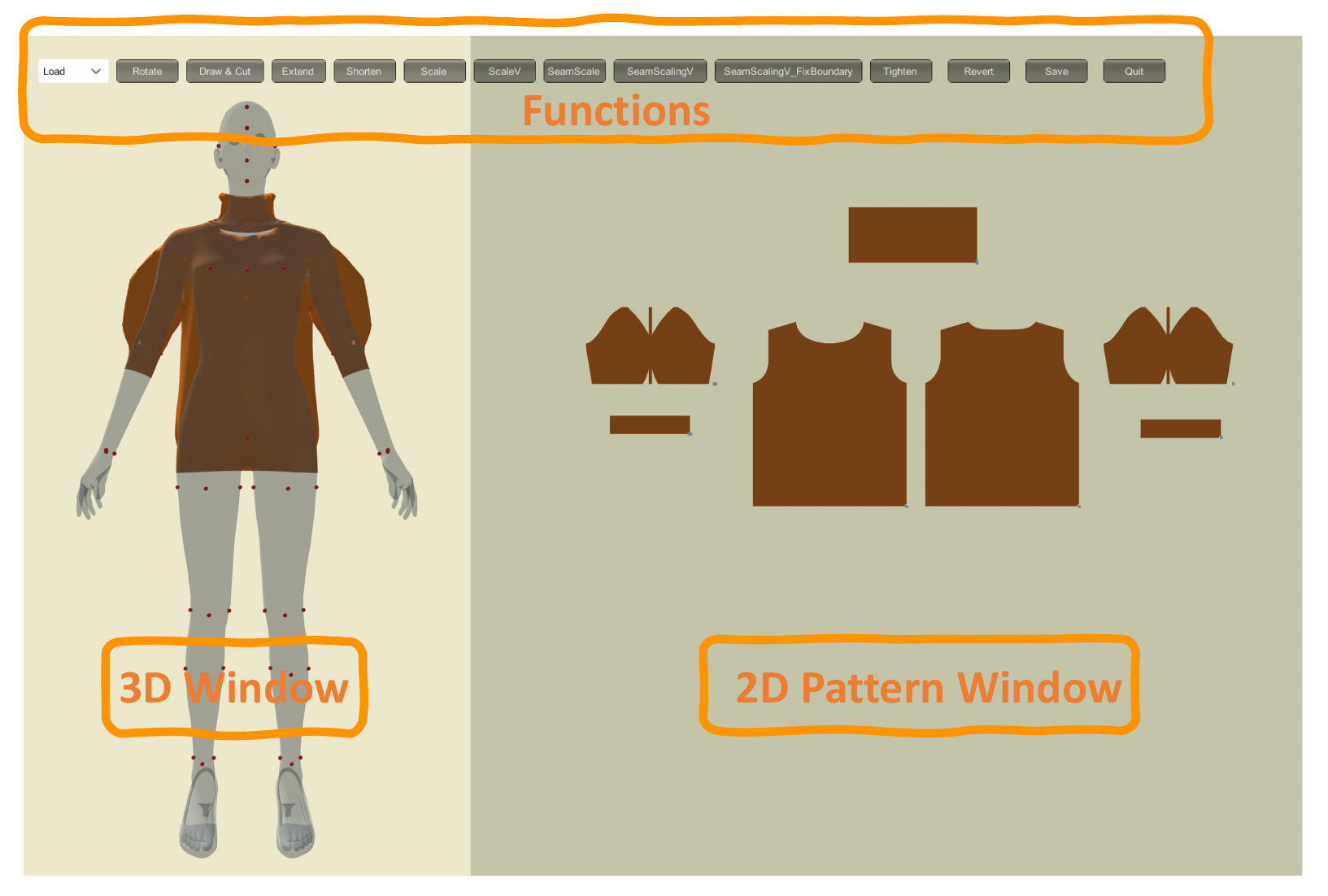}
    \vspace{-0.5cm}
    \caption{A screen snapshot of the system. The left (beige) is the 3D window displaying the 3D garment on a human body. Red dots indicate key feature points on the surface of the human body, \eg, front neck point, and busty points. The right is the 2D pattern window showing the sewing pattern. The top shows the functions of the system.}  
    \label{interface}
\end{figure}

The system has two windows: the 3D window and the 2D pattern window
for displaying the 3D garment and 2D sewing pattern, respectively (see Figure~\ref{interface}). The user can edit the 3D garment using editing operations provided by the system in the 3D window. In the 2D pattern window, the user can freely move the panel to adjust the pattern's layout by clicking. The top shows the system's editing functions. Since interacting with these editing functions is straightforward and hence not described here. Please see the accompanying video for details. In the following section, we detail the supported operations and illustrate them with multiple examples.

\begin{figure*}[ht]
 \centering
  \includegraphics[width=0.99\textwidth]{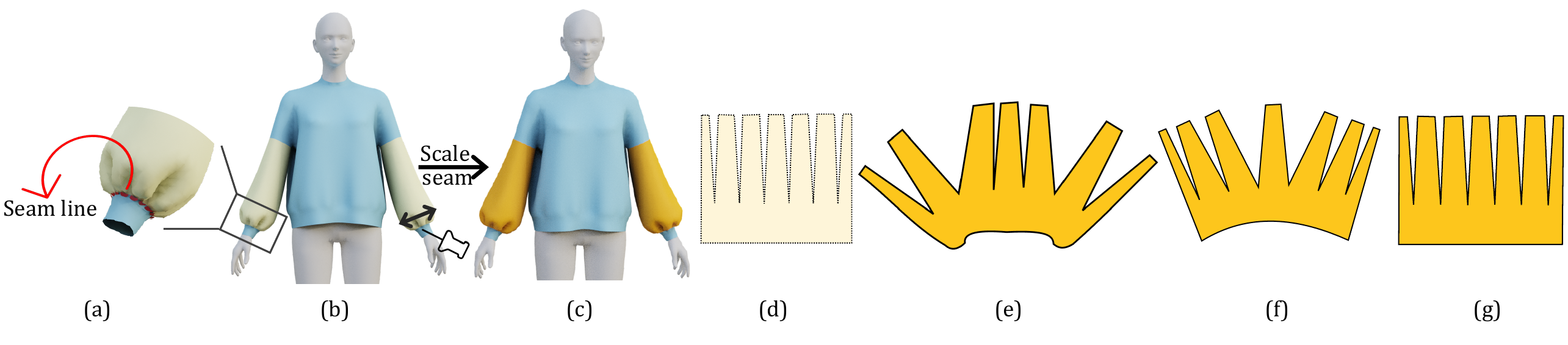}
  \vspace{-0.3cm}
  \caption{Scale the internal sleeve part by editing the seam line. \textmd{(a) The red stitches visualize the seam line of the lantern sleeve. (b) The user edits the seam line to scale the internal sleeve (soft yellow) perpendicular to the human body whilst fixing the bottom part around the wrist. (c) The scaled internal sleeve (hard yellow). (d) The original sleeve panel. (e) The updated panel by Igarashi \etal$^{4}$ (f, g) The updated panel by our method without/with the as-similar-as-possible constraint, respectively.  Compared with (g), Igarashi \etal$^{4}$ (e) has a slightly larger pattern with the severely shrunk lower boundary. This is because it directly flattens the deformed 3D triangles, which are deformed due to stitching force, gravity, and \emph{etc} in the simulation. (f) has the right size but its curved boundary damages the original design. With our as-similar-as-possible constraint, (g) has the right size whilst maintaining the original panel's boundary design.}}
  \label{seam}
\end{figure*}

\subsection{Editing Operations}
\begin{figure}[h!]
    \centering
    \includegraphics[width=1.0\linewidth]{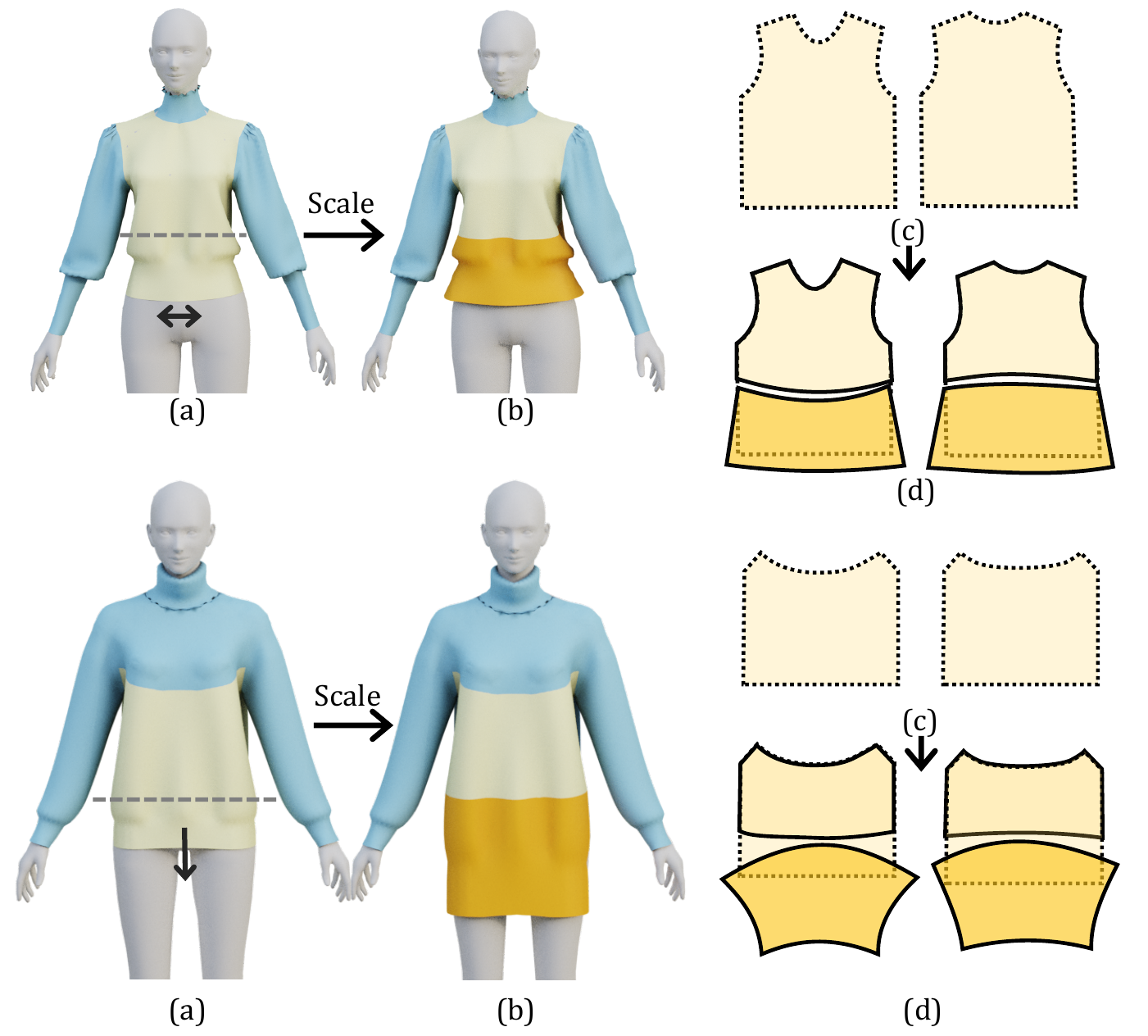}
    \vspace{-0.5cm}
    \caption{The user scales the bottom part of the garment perpendicular to the human skeleton making it looser (top row), along with the human skeleton making it longer (bottom row). (a) The original garment. Soft yellow indicates the panels affected by the user edits. (b) The updated garment geometry. Dark yellow indicates the parts selected by the user, being customised. (c) The original panels. (d) The updated panels (Dash lines illustrate the original panel shape).}
    
    \label{scale}
\end{figure}

\begin{figure}[h!]
    \centering
    \includegraphics[width=1.0\linewidth]{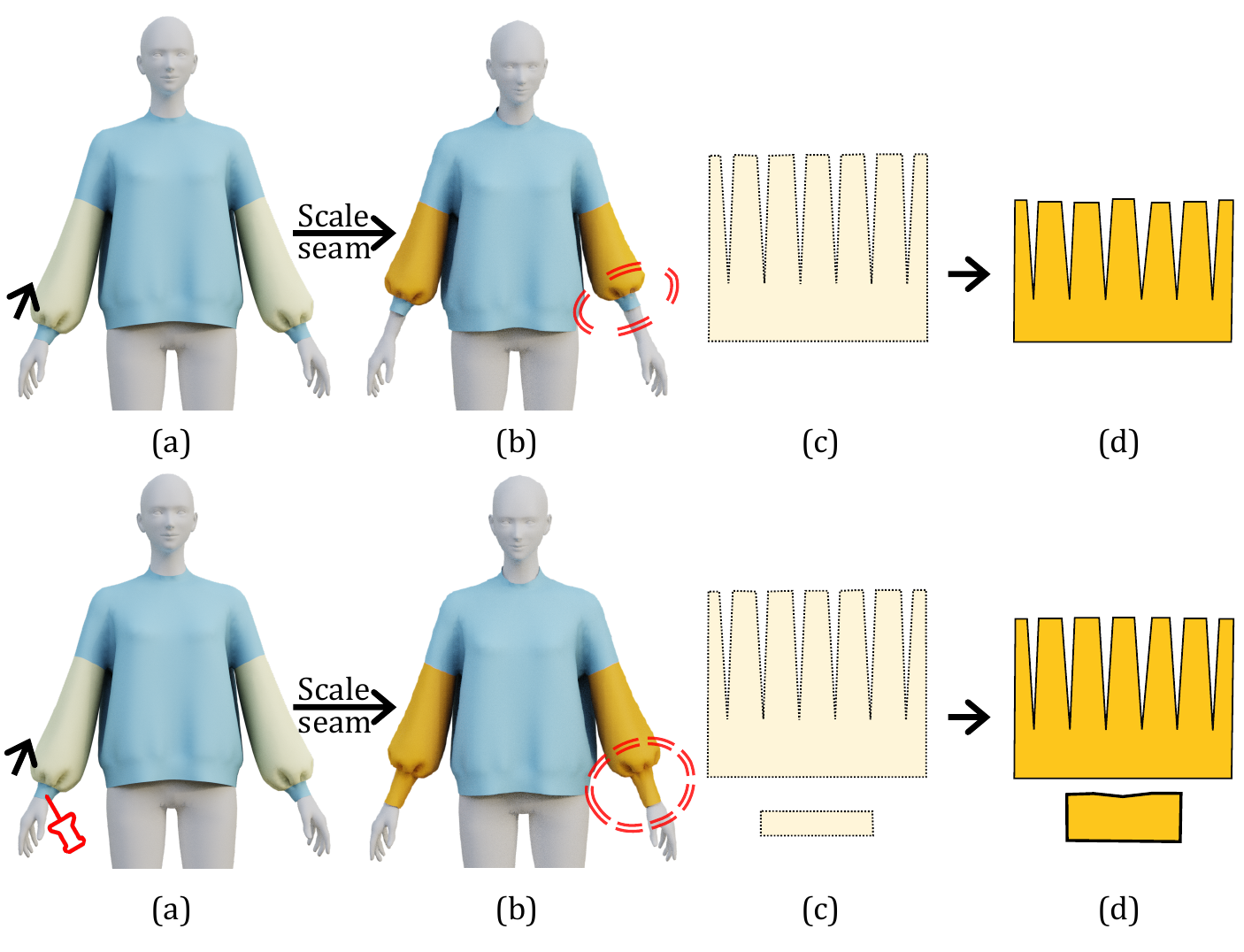}
    \vspace{-0.5cm}
    \caption{Scale the internal sleeve part along the body by editing the seam line. \textmd{Top row: The user moves the seam line upwards making the internal sleeve shorter. Bottom row: The user moves the seam line upwards whilst fixing the lower boundary of the bottom sleeve. This leads to a shorter internal sleeve but a longer bottom sleeve. The red circle in (b) highlights the difference. (a) The original garment. Soft yellow indicates the panels affected by the user edits. (b) The updated garment geometry. Dark yellow indicates the parts selected by the user and being customised. (c) The original panels. (d) The updated panels.}}
    
    \label{scaleseam}
\end{figure}

Starting with the input 3D garment and the corresponding pattern, we implement a set of simple and commonly used alteration operations that allow the user to explore the complex redesign space:

\noindent\textbf{Scale}: The user can select a part on the 3D garment and scale either along (see Figure~\ref{scale} top row) or perpendicular (see Figure~\ref{scale} bottom row) to the human skeleton direction by dragging. Furthermore, our system also supports the scaling of the internal garment part leveraging the seam line. In the garment literature, a seam line refers to the line or path created by joining two or more pieces of fabric together using stitches (see Figure~\ref{seam} (a)). Our system utilizes the seam line to allow the user to scale the internal garment part either along (see Figure~\ref{seam}) or perpendicular (see Figure~\ref{scaleseam}) to the human body. In such instances, after thorough discussions with the designers, we opt to update the whole sewing panel using our proposed flattening method with the as-similar-as-possible constraint to preserve the original panel's boundary shape. The rationale behind this is the user edit will affect the entire sewing panel equally, thus it is optimal to maintain the original panel's boundary design as closely as possible.

\noindent\textbf{Cut}: We allow the user to sketch on the 3D garment to cut the garment into the desired shape. By default the user sketches from one viewpoint, and we will automatically form a closed loop as the cutting line on the 3D garment to cut both sides. We also enable the user to only cut one side if the user specifies. Then we re-triangulate the meshes affected by the new cutting line and update the sewing pattern by directly transferring the barycentric coordinate of the new vertex into the local coordinate system of the corresponding 2D triangle. For the detailed algorithm, we refer the reader to the Teddy system.$^3$

\begin{figure}[h!]
    \centering
    \includegraphics[width=0.85\linewidth]{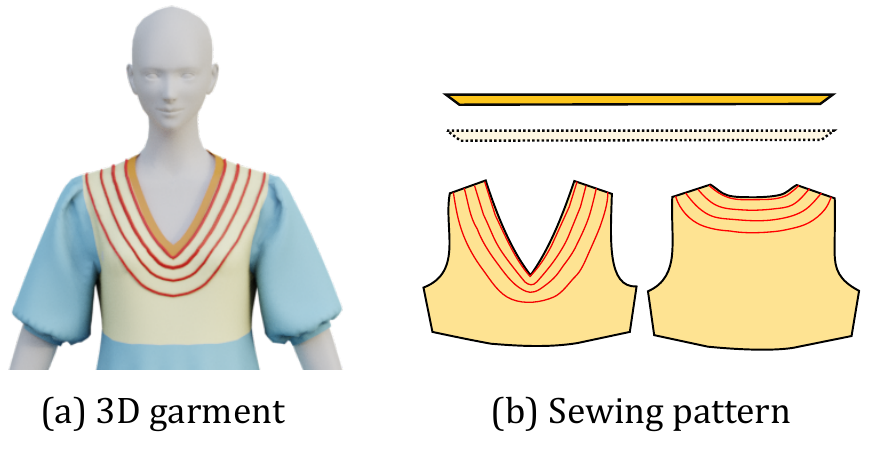}
    \vspace{-0.3cm}
    \caption{Shorten example. The user iteratively shortens the garment four times to explore various collar designs. Red curves on (a) and (b) indicate the cutting lines on the 3D garment and sewing pattern, respectively.}
    
    \label{shorten_steps}
\end{figure}

\noindent\textbf{Shorten}: The user can drag the boundary to the desired position to shorten the garment (see Figure~\ref{teaser} (b) Step 3 and Figure~\ref{shorten_steps}). In detail, we compute the iso-line on the garment surface mesh where the distance to the boundary is the user-specified shortened distance. We take the computed iso-line as the cutting line on the 3D garment and utilize the Cut function to shorten the garment.

\begin{figure}[h!]
    \centering
       \vspace{0.1cm}
    \includegraphics[width=0.85\linewidth]{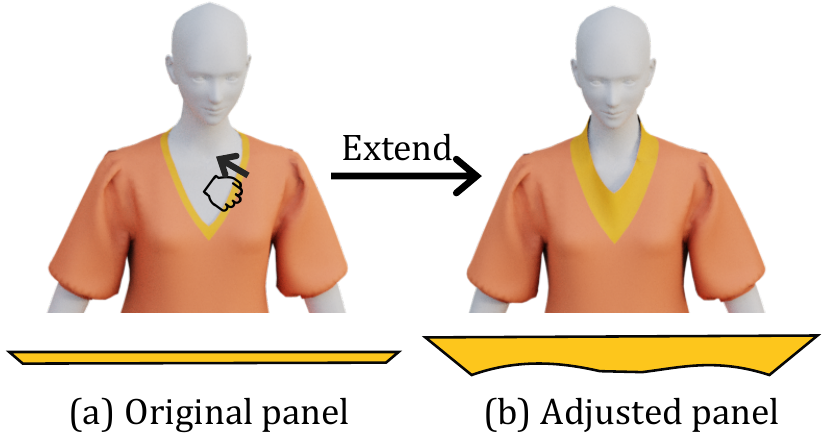}
    \vspace{-0.3cm}
    \caption{The user extends the V-shape neck collar.}
    
    \label{extend}
\end{figure}

\noindent\textbf{Extend}: The user can also extend the garment by dragging the boundary to the desired position (see Figure~\ref{teaser} (b) Step 2 and Figure~\ref{extend}). Following the observation from Brouet \etal$^2$, people tend to preserve the slope and tangent plane orientations across the garment surface when transferring the garment. We follow the same principle by appending the triangle faces which share the same surface norm with the connected triangle faces. Body-garment collisions sometimes occur when extending. We resolve it by pushing the vertex towards the norm direction of the nearest triangle on the body surface. Finally, we update the sewing pattern in the same way as that of the Cut function.  

\noindent\textbf{F5 Tighten/Loosen}: We also allow the user to tighten/loosen the garment by directly over-sketching the silhouette of the 3D garment. The system deforms the garment to meet the silhouette and updates the pattern with our proposed flattening method. For the detailed 3D deformation algorithm, we refer the reader to the Nealen \etal$^{16}$

Those editing operations allow the user to alter garments and showcase the capabilities of our proposed pattern adjustment method, but more operations could be added to enhance the redesign capacity such as adding folds and darts.

\section{CONCLUSION}
We present a pattern adjustment method that aims to preserve the embedded design for garment alteration. Besides, we develop a set of editing operations to support alteration and showcase the capability of our adjustment method and functions via several examples throughout the paper. The editing operations introduced in this paper are not exhaustive, the the core algorithm, "scale-preserving flattening" is general and can be applied to other 3D modelling operations.

\section{ACKNOWLEDGMENT}
This work is supported by SHIMA SEIKI MFG., LTD. This work is also partially supported by JST CREST (JPMJCR17A1).

\begin{IEEEbiography}{Anran Qi}{\,}is a Project Assistant Professor at the Graduate School of Information Science and Technology at The University of Tokyo. Her research focuses on garment design and human sketch understanding from graphics, machine learning, and human-computer interaction perspectives. She received a Ph.D from the Centre for Vision, Speech and Signal Processing (CVSSP), University of Surrey in 2021. Contact her at annranqi@gmail.com.
\end{IEEEbiography}
\begin{IEEEbiography}{Takeo Igarashi}{\,}is a Professor of Computer Science Department at The University of Tokyo. His research interest is in user interfaces and interactive computer graphics in general. He received a Ph.D. from the Department of Information Engineering at The University of Tokyo in 2000. He received the SIGGRAPH Significant New Researcher Award, JSPS Award, CHI Academy Award, and the Asia Graphics 2020 Outstanding Technical Contributions Award. Contact him at takeo@acm.org.
\end{IEEEbiography}

\end{document}